\documentclass[11pt]{evolution}    

\usepackage{natbib}
\usepackage{color}

\usepackage{epsfig,amsmath,amsfonts,amssymb,cases,subfig,ulem,wasysym} 
\newcommand{\be}{\begin{equation}}
\newcommand{\ee}{\end{equation}}
\newcommand{\bea}{\begin{eqnarray}}
\newcommand{\eea}{\end{eqnarray}}
\newcommand{\bml}{\begin{mathletters}}
\newcommand{\eml}{\end{mathletters}}

\newcommand{\tc}{\textcolor}

\renewcommand{\citep}[1]{(\citealt{#1})}

\raggedright

\begin{document}

\title{Fixation probability of rare nonmutator and evolution of mutation rates}

\author{Ananthu James %
       \email{Ananthu James - ananthujms@jncasr.ac.in}%
      and
         Kavita Jain$^{*}$%
         \email{Kavita Jain - jain@jncasr.ac.in}}

\address{Theoretical Sciences Unit,
Jawaharlal Nehru Centre for Advanced Scientific Research, Jakkur P.O.,
Bangalore 560064, India
}

\maketitle
Running Title : Evolution of mutation rates

\vspace{5mm}
{\bf Contact Information (for all authors)} \\

Ananthu James

{\bf postal address}: Theoretical Sciences Unit,
Jawaharlal Nehru Centre for Advanced Scientific Research, Jakkur P.O.,
Bangalore 560064, India

{\bf work telephone number}: +91-80-22082967

{\bf E-mail}: ananthujms@jncasr.ac.in \\

Kavita Jain 

{\bf postal address}: Theoretical Sciences Unit,
Jawaharlal Nehru Centre for Advanced Scientific Research, Jakkur P.O.,
Bangalore 560064, India

{\bf work telephone number}: +91-80-22082948

{\bf E-mail}: jain@jncasr.ac.in

\clearpage

\begin{abstract}
Although mutations drive the evolutionary process, the rates
  at which the mutations occur are themselves subject to evolutionary
  forces. Our purpose here is to understand the role of selection and
  random genetic drift in the evolution of mutation rates,  
\tc{black}{and we address this question in asexual populations at mutation-selection equilibrium neglecting selective sweeps. 
Using a multitype branching
process, we calculate the fixation probability of a
rare nonmutator in a large asexual population of mutators, and find that  a nonmutator is more likely to fix when the deleterious
mutation rate of the mutator population is high. Compensatory mutations in the mutator population are found to
  decrease the fixation probability of a nonmutator when the selection
  coefficient is large. But, surprisingly, the fixation probability changes nonmonotonically with increasing compensatory mutation rate when the selection is mild.}  
Using these results for the fixation probability and a drift-barrier argument, we find a novel relationship between the mutation rates and the population size. We also discuss the time to fix the nonmutator in an adapted
population of asexual mutators, and compare our 
results with experiments.

\vspace{1cm}

\end{abstract}

\clearpage

%=========================================================================
%=========================================================================
%INTRODUCTION
%=========================================================================
%=========================================================================
Because most mutations are deleterious, the mutation rate can
not be too high; in fact, in an infinitely large population, for a
broad class of fitness functions, an error
threshold has been shown to exist above  which the deleterious effects
of mutation can not be compensated by selection  \citep{Eigen:1971,Jain:2007b}. The mutation rate is not
zero either \citep{Baer:2007}, and it has been argued that the
stochastic fluctuations in a finite population limit the evolution of 
mutation rates below a certain level \tc{black}{since in small enough
populations, the advantage gained by lowering the mutation rate can
not compensate the effect of random genetic drift
\citep{Lynch:2010b}.} 
Empirical
data for organisms with widely different effective population size shows a
negative correlation between the deleterious mutation rate 
and the population size \citep{Sung:2012}, and some quantitative insight into this 
relationship has been obtained by treating all deleterious mutations
to be  lethal
\citep{Lynch:2011}. However, this is clearly an extreme scenario, and
it is important to ask how the deleterious mutation rate evolves when mutations are only weakly deleterious.

Many theoretical and experimental investigations have also
shown that in an adapting asexual population, a mutator allele causing
a higher mutation rate than that of the nonmutator can get fixed (see
a recent review by \citet{Raynes:2014}). \tc{black}{Since the mutators produce not only deleterious
but also beneficial mutations at a higher rate than the nonmutators,
the mutator allele can hitchhike to fixation with favorable mutations
\citep{Smith:1974,Taddei:1997}.}  
However, once the population has reached a high fitness level, high mutation 
rates are detrimental since most mutations will now be deleterious,
and in such a situation, the mutation rate is expected to decrease \citep{Liberman:1986a}. Indeed, in some experiments 
\citep{Trobner:1984,Notleymcrobb:2002,Mcdonald:2012,Turrientes:2013,Wielgoss:2013},   
the mutation rate of an adapted population carrying a mutator allele
has been seen to decrease and the time to fixation has been measured, 
but a theoretical understanding of this time scale is missing. 

To address the issues discussed above, 
we study the fate of a rare nonmutator in a large asexual population of mutators 
using a multitype branching process \citep{Patwa:2008}. An important
difference between the previous works on mutator hitchhiking
\citep{Taddei:1997,Andre:2006,Wylie:2009,Desai:2011} and our study is
that here the mutator population
is assumed to be at  mutation-selection equilibrium and is therefore
not under positive selection. However, compensatory mutations that alleviate the effect
of deleterious mutations are included in our model. 
We find that when only deleterious mutations are present, a 
nonmutator can get fixed with a probability that increases with
the deleterious mutation rate of the mutator. 
Compensatory mutations in the mutator
population are expected to decrease the fixation probability of the
nonmutator, 
and we find that this intuition is indeed correct when deleterious
mutations in the mutator are effectively lethal. But,  
surprisingly, when the deleterious mutations are mildly harmful, the
fixation probability is found to initially increase and then decrease
as the rate of compensatory mutations increases. Our study thus identifies the conditions
under which the spread of nonmutators is suppressed in the absence of
positive selection, and complements earlier works in which a
mutator hitchhikes with  beneficial mutations to fixation  \citep{Taddei:1997,Andre:2006,Wylie:2009,Desai:2011}.

Using our results for the fixation probability and a 
 drift-barrier argument which states that the advantage offered by a
 decrease in the deleterious mutation rate is
 limited by random genetic drift in a finite population \citep{Lynch:2010b}, we find that the deleterious mutation  rate decreases
with increasing population size in accordance with experimental data
\citep{Sung:2012}. However, unlike previous theoretical work that treats the
deleterious mutations to be effectively lethal \citep{Lynch:2011}, 
here we consider
both strongly and weakly deleterious mutations, and not only
reproduce the result in \citet{Lynch:2011}, but also find a new
scaling law in the latter case. 
We also use the results for the fixation probability to find the time
to lower the mutation rate in an adapted population of mutators, and 
compare our theoretical results with recent experiments \citep{Mcdonald:2012,Wielgoss:2013}. 

%===========================================================================
%===========================================================================
%MODELS 
%===========================================================================
%===========================================================================
\section*{Model and methods}

We consider an asexual population in which the fitness of an
individual with $k$ deleterious mutations is given by $W(k) =(1-s)^k$,
where the selection coefficient $0 < s < 1$. A deleterious
mutation is allowed to occur at a rate $U_d$ and a beneficial one at a 
rate $U_b < U_d$. We are interested in the fate of a
nonmutator that arises in this population and whose total 
mutation rate is smaller than that of the mutator. 
In a sufficiently large population of mutators in which stochastic
fluctuations {due to genetic drift} may 
be ignored, this can be addressed using a branching process
\citep{Patwa:2008}, as described below. 

\begin{figure}
\begin{center} 
\includegraphics[width=0.69\textwidth,angle=-90]{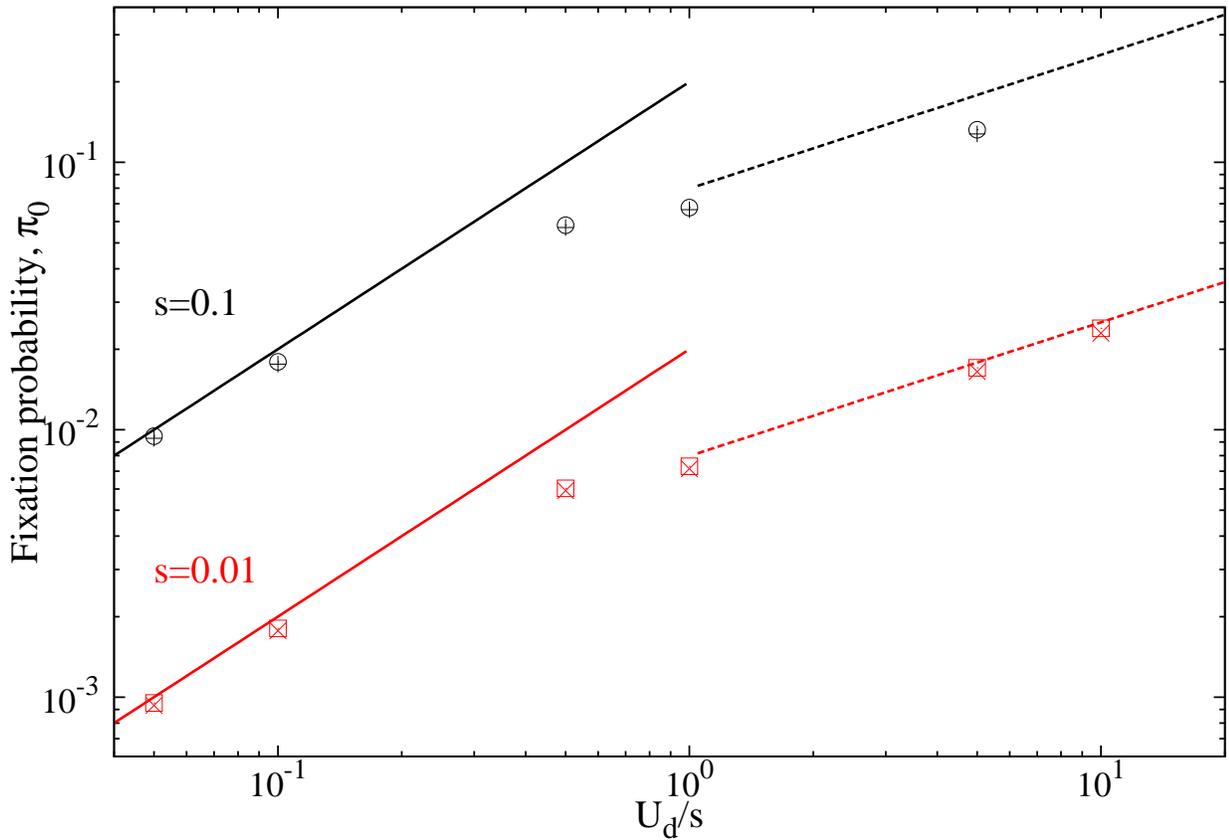}
\caption{Dependence of the fixation probability obtained using a
  multitype branching process on the deleterious mutation
  rate $U_d$ for two values of the selection coefficient $s$ and
  compensatory mutation rate $U_b=0$. The points 
  are obtained by numerically solving (\ref{Piss2}) when the mutation
  rate of the nonmutator is zero  ($\Circle,\Square$), and the
  stationary state solution of (\ref{Piktfull}) 
  when the nonmutator's mutation rate is $50$ times
  lower than that of the mutator ($+, \times$). The lines show
  the analytical result (\ref{Pi0analy}).} 
\label{piVd}
\end{center} 
\end{figure}

The fixation probability $\pi(k,t)$ of a single copy of a 
nonmutator allele with fitness 
$W(k)$ present at generation $t$ changes according to  \citep{Johnson:2002}
\be
1- \pi(k,t) = \exp \left[-\frac{W(k)}{{\overline W}(t)}
  \sum_{k'} M(k \to k') \pi(k',t+1) \right], 
  \label{Piktfull}
\ee
where ${\overline W}(t)= \sum_{k=0}^\infty W(k) p(k,t)$ is the average
fitness of the mutator population and $p(k,t)$ is the mutator frequency. The above equation expresses the fact that a single copy of the rare
allele in the fitness class $k$ whose offspring distribution is
  Poisson with mean $W(k)/{\overline W}(t)$ will be lost eventually if
  each of  its
offspring, which may undergo mutations with probability $M(k \to k')$, do not
survive. Here we consider strong mutators  
whose mutation rate is much higher than that of the nonmutator
\citep{Sniegowski:1997,Oliver:2000}, and therefore neglect the 
mutation rate of the latter in most of the following discussion (however, see Fig.~\ref{piVd}). 
We also assume that the mutator 
population is at mutation-selection equilibrium as is likely to be 
the case in large populations that have been evolving for a long time
in a constant environment. As a result, the probability $\pi(k,t)$ becomes time-independent. 
These considerations lead to a relatively 
simpler, but still highly nonlinear equation given by 
\be
1- \pi(k) = \exp \left[-\frac{W(k) ~\pi(k) }{{\overline W}} \right]~.
\label{Piss2}
\ee
The above expression, of course, reduces to the well known
single-locus equation \citep{Fisher:1922,Haldane:1927b} 
when the nonmutator can be present in only one genetic background, but  
here we are dealing with a multitype branching process since a
nonmutator can arise in any fitness class. 

The total fixation probability is obtained on
summing over all genetic backgrounds \citep{Johnson:2002},
\be
\pi_{\textrm{tot}}= \sum_{k=0}^\infty p(k)~\pi(k)~,
\label{Ptot}
\ee 
where the probability that a nonmutator arises in a
  background of $k$ deleterious mutations is given by the mutator
  frequency $p(k)$ in the stationary state.
  
Although the steady state frequency $p(k)$ in the absence of compensatory mutations that mitigate the effect of deleterious mutations is known exactly \citep{Kimura:1966,Haigh:1978}, the corresponding
solution with nonzero $U_b$ is not known. We therefore compute the
mutator frequency numerically for nonzero $U_b$ using
(\ref{Qdyn}) given in Appendix~\ref{apdetda},  
 and use these results in (\ref{Piss2}) to find the
fixation probability for arbitrary $U_b$. 
To make analytical progress, we use a perturbation theory in
  which the effect of the small dimensionless parameter $U_b/s$ can be studied by
  expanding the quantities of interest in a power series in $U_b/s$, and 
  write 
\be
\pi(k)= \sum_{n=0}^\infty \left( \frac{U_b}{s} \right)^n \pi_n(k)~,~ p(k)=
\sum_{n=0}^\infty \left( \frac{U_b}{s} \right)^n {p}_n(k)~.
\label{pert}
\ee
The terms $\pi_0(k)$ and $p_0(k)$ corresponding to $n=0$ in the above expansion give the
results in the absence of compensatory mutations, and in
Appendix~\ref{apdetda}, we calculate the stationary state fraction $p(k)$ to
linear order in $U_b/s$.

%===========================================================================
%RESULTS
%===========================================================================

\section*{Results}

%------------------------------------------------------------------------
\subsection*{Fixation probability}

{\bf In the absence of compensatory mutations:} We first consider the
case when $U_b=0$. Taking the logarithm
on both sides of (\ref{Piss2}), and expanding the left hand side (LHS) 
up to $\pi_0^2(k)$, we find that either $\pi_0(k)=0$, or 
\bea
\pi_0(k) = 2 \left(\frac{W(k)}{\overline{W_0}}-1 \right)  \approx 2 s ({\bar k}_0-k)~,
\label{Pi01}
\eea
where  the average fitness $\overline{W_0}=e^{-U_d}$ and the average number of
deleterious mutations ${\bar k}_0=U_d/s$
\citep{Kimura:1966,Haigh:1978}. The last expression on the right hand
side (RHS) of (\ref{Pi01}) is obtained by expanding the exponentials
as the parameters $U_d$ and $s$ are small.  
Since the fixation probability must
not be negative, the expression (\ref{Pi01}) is
valid when $k < \lfloor {\bar k}_0\rfloor$, and the solution
$\pi_0(k)=0$ holds otherwise. Here $\lfloor x \rfloor$ denotes the largest
integer less than or equal to $x$. More generally, a nonmutator can
get fixed if its fitness $W(k) \approx e^{-s k}$ is larger than
the average fitness $e^{-s {\bar k}}$  of the mutator population, or
$k < \lfloor{\bar k}\rfloor$, ${\bar k}$ being the average number of
deleterious mutations \citep{Johnson:2002}.

Equation (\ref{Pi01}) shows that the fixation probability $\pi_0(k)$ decreases as the
number of deleterious mutations increase, as one would intuitively
expect. However, the probability
 $p_0(k)$ that a nonmutator would arise in a background with $k <
  {\bar k_0}$ deleterious mutations increases. 
On summing over the backgrounds in which a nonmutator can
  arise, as explained in Appendix~\ref{apdet0}, we find that the total fixation probability falls in two
distinct regimes defined by whether $U_d$ is below or above $s$:
\be
{\pi_0=\sum_{k=0}^{\lfloor {\bar k}_0 \rfloor} \pi_0(k) p_0(k)}=
\begin{cases}
2 U_d  ~,~U_d \ll s \\
\sqrt{\frac{2 s U_d}{\pi}}  ~,~U_d \gg s~.
\end{cases}
\label{Pi0analy}
\ee
For ${\bar k}_0 \ll 1$, since a mutation is costly, it can be
  treated as effectively lethal \citep{Johnson:1999b}. 
In this situation, the advantage conferred by the
  nonmutator is simply given by $1-e^{-{U_d}} \approx U_d$ and the
  classical result for the single locus problem gives the fixation
  probability to be $2 U_d$ \citep{Fisher:1922,Haldane:1927b}. 
For ${\bar k}_0 \gg 1$, the total fixation
probability apparently receives contribution from ${\bar k}_0$ genetic
backgrounds, but merely $\sqrt{{\bar k}_0}$ genetic backgrounds are actually 
relevant because the Poisson-distributed frequency $p_0(k)$
has a substantial weight for fitness classes that lie within a width
$\sqrt{{\bar k}_0}$ of the mean (also, see Appendix~\ref{apdet0}). 
%which gives $\pi_0 \sim 2 s
%\int_{{\bar k}_0-\sqrt{{\bar k}_0}}^{{\bar k}_0} dk ({\bar k}_0-k)= 2
%s \sqrt{\bar{k}_0}$, as obtained above. 
Equation (\ref{Pi0analy}) shows that for fixed $s$, the nonmutator is more
 likely to be fixed when $U_d$ is large. But, for a given $U_d$, the
 fixation probability initially increases with the selection
 coefficient and then saturates 
 to $2 U_d$. In Fig.~\ref{piVd}, the analytical results above are
compared with those obtained by numerically iterating 
(\ref{Piss2}) and (\ref{Piktfull}) when the mutation rate of the nonmutator is zero and
$U_d/50$ respectively, and we see a good agreement in both cases.

\begin{figure}
\begin{center} 
\includegraphics[width=0.7\textwidth,angle=-90]{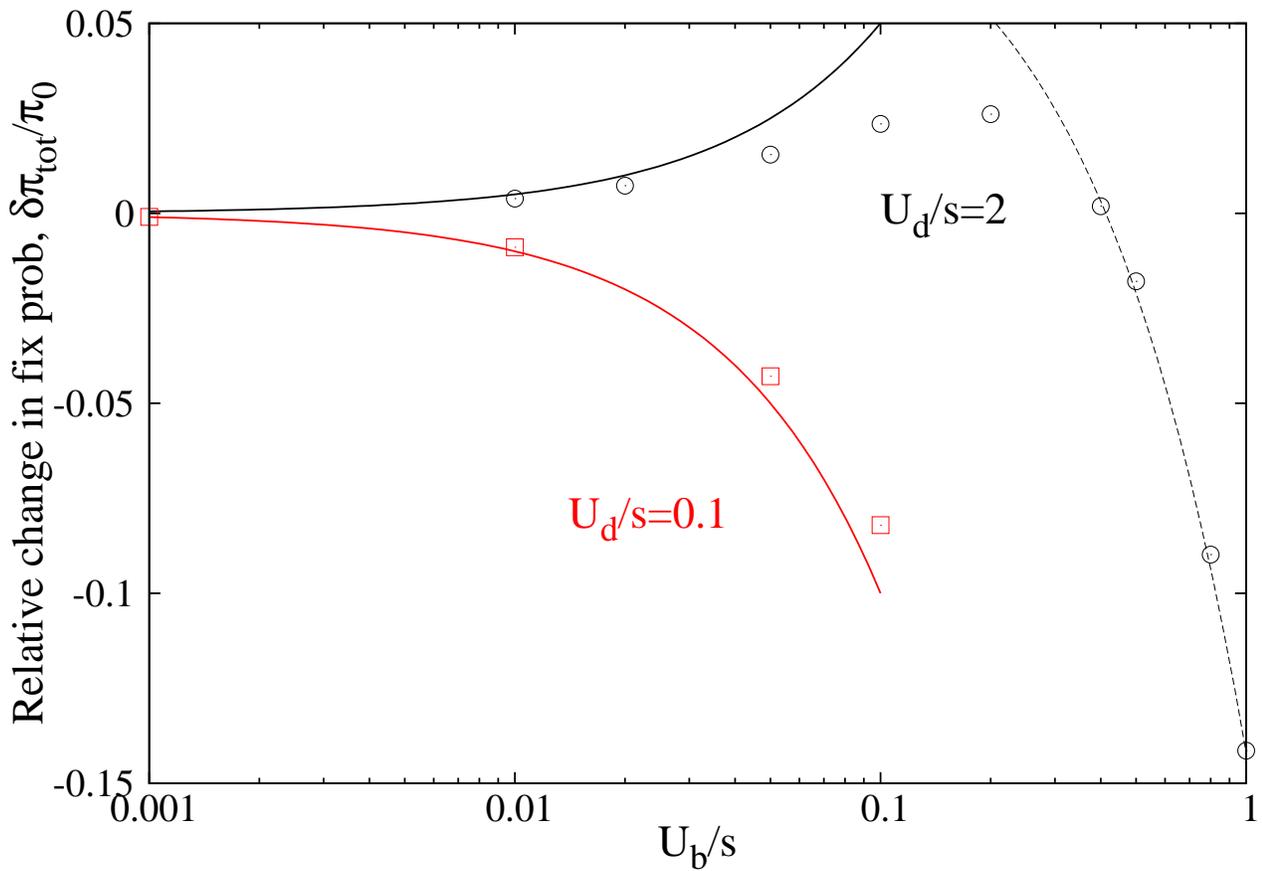}
\caption{Dependence of the fixation probability obtained using a
  multitype branching process on the 
  compensatory mutation rate $U_b$ for two values of $U_d/s$. The points
  show the numerical solution of (\ref{Piss2}) and the lines show the
  analytical results (\ref{ssben}) and (\ref{wsben}). The
  broken curve for $U_b/s > 0.1$ is a linear fit, $0.1-0.24 U_b/s$, to the numerical data. For $U_d/s=0.1$, the
  ratio $U_b/s$ is also 
  below $0.1$ since $U_b$ is assumed to be smaller than $U_d$. } 
\label{piVb}
\end{center} 
\end{figure}

{\bf Including compensatory mutations:} We now study how compensatory 
mutations in the mutator population affect the fixation probability of the
nonmutator. Figure~\ref{piVb} shows that when ${\bar k}_0 \ll 1$, the fixation probability decreases with $U_b$, but for 
${\bar k}_0 \gg1$, it changes {\it nonmonotonically}: it first
increases and then decreases with increasing $U_b$. To understand this behavior, consider the change $\delta
  \pi_{\textrm{tot}}=\pi_{\textrm{tot}}-\pi_0$ in the fixation
  probability due to compensatory mutations which is simply given by 
\be
\delta \pi_{\textrm{tot}}= \sum_{k=0}^{\lfloor \bar k \rfloor} p_0(k) \delta
\pi(k)+ \pi_0(k) \delta p(k)+ \delta p(k) \delta \pi(k)~.
\label{change}
\ee
When $U_b$ is nonzero, the 
  change in the fixation probability $\delta \pi(k)=\pi(k)-\pi_0(k)$
  and the  mutator frequency
  $\delta p(k)=p(k)-p_0(k)$ behave in a qualitatively different manner. With 
  increasing $U_b$, the average {\it fitness} of
  the mutator population increases which, by virtue of (\ref{Piss2}),  decreases the fixation probability of the nonmutator, {\it i.e.} $\delta
  \pi(k) < 0$.  
However, since the {\it frequency} of individuals 
with less deleterious mutations increases when $U_b$ is nonzero, the
change in the mutator fraction $\delta p(k) > 0$. 
Thus the change in the total fixation 
probability given by (\ref{change}) receives both positive and negative contributions, and it
is not obvious which one of these factors would have a larger effect. 

\begin{figure}
\begin{center} 
\includegraphics[width=0.7\textwidth,angle=-90]{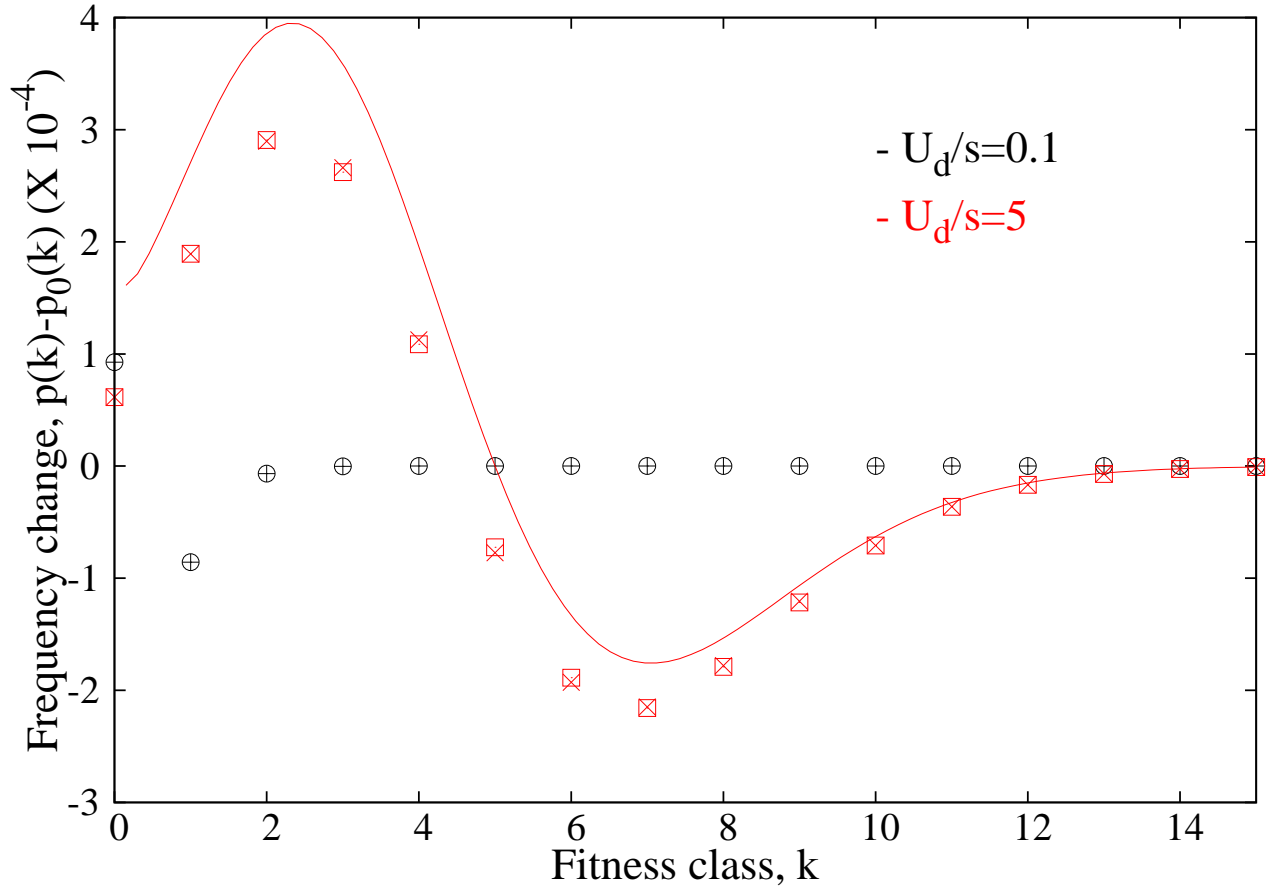}
\caption{Change in the mutator frequency when compensatory mutations
  are included, $\delta p(k)=p(k)-p_0(k)$ for $U_b=10^{-4}$. The points
  ($\Circle,\Square$)  are obtained by numerically iterating
  (\ref{Qdyn}) and ($+,\times$) show the 
  perturbation theory result (\ref{Q1corr}), and we observe a good
  agreement. The simple expression 
  (\ref{Q1simple}) for large $U_d/s$ is also shown (lines).} 
\label{freqVb}
\end{center} 
\end{figure}

To address this question, we calculate the fixation probability for small $U_b/s$ as described below. 
Substituting (\ref{pert}) in the expression (\ref{change}) for $\delta \pi_{\textrm{tot}}$,
and neglecting terms of order $(U_b/s)^2$ and higher, we find that $\delta \pi_{\textrm{tot}} \approx (U_b/s) \pi_1$, where
\be
\pi_1 =\sum_{k=0}^{{\lfloor {\bar k}_0 \rfloor}} p_0(k) \pi_1(k)+ p_1(k) \pi_0(k) ~.
\label{Pi1expr}
\ee
The 
contribution $\pi_1(k)$ is calculated in Appendix~\ref{apdetdc}, and we find that
\bea
\pi_1(k) \approx  -2 s {\bar k}_0 (1-\pi_0(k))~,~k < \lfloor {\bar k}_0
\rfloor~, 
\label{Pi1simple}
\eea
which is negative, as expected. 
An expression for the fraction $p_1(k)$ is obtained in Appendix~\ref{apdetda}, and its behavior
is shown in Fig.~\ref{freqVb} for small and large ${\bar k}_0$. 
For small ${\bar k}_0$, the frequency $p_0(k)$ is close to one in the
zeroth fitness class and zero elsewhere. But the correction $p_1(k)$ is negligible in
all the fitness classes. 
For large ${\bar k}_0$, the contribution $p_1(k)$ is significantly
different from zero in many fitness
classes and can be approximated by 
\be
p_1(k)= {\bar k}_0 p_0(k) ~\ln \left( \frac{{\bar k}_0}{k} \right) ~,~k
\gg 1~.
\label{Q1simple}
\ee
Thus, as claimed above, the fraction $p_1(k)$ is
positive for $k < {\bar k}_0$ and negative for $k > {\bar k}_0$ (also,
see Fig.~\ref{freqVb}).

When $U_d \ll s$, as already mentioned, the fraction $p_1(k)$ is
negligible in all the fitness
classes and $p_0(0) \approx 1$. Using these results in (\ref{Pi1expr}) and (\ref{Pi1simple}), we get
$\pi_1=-2 s {\bar k}_0$, and thus
\be
\frac{\delta \pi_{\textrm{tot}}}{\pi_0}= -\frac{U_b}{s} ~,~ U_b < U_d < s.
\label{ssben}
\ee
This reduction in the fixation probability of the nonmutator when $U_b$
is nonzero is expected since the effect of compensatory 
mutation is to restore the mutators that have suffered lethal mutation
to the zeroth mutation class thus enabling them to offer competition to the nonmutators.

When $U_d \gg s$, as shown in  Appendix~\ref{apdetdc}, we can obtain a quantitative estimate of the initial
increase in $\delta \pi_{\textrm{tot}}$ by calculating the sum on the RHS of
(\ref{Pi1expr}) to obtain  (\ref{Vbres}), and thence 
\be
\frac{\delta \pi_{\textrm{tot}}}{\pi_0}=\frac{U_b}{2 s} ~,~U_b < s < U_d~.
\label{wsben}
\ee
Thus we find that for small $U_b$, the increase of the mutation
frequency in fitness classes with fewer deleterious mutations dominates the increase in the mutator
fitness resulting in positive $\delta \pi_{\textrm{tot}}$.  However, for large $U_b$, the net change in the fixation probability is negative since the last term in
the summand of (\ref{change}), which is also negative, 
 enters the picture. 
Since the maximum in $\delta \pi_{\textrm{tot}}$ occurs at large $U_b/s$, the perturbation
theory described here can not capture the eventual decrease in this parameter 
regime. A quantitative comparison of the results obtained by numerically
solving (\ref{Piss2}) and (\ref{Qdyn}) for arbitrary $U_b$ with the analytical
results (\ref{ssben}) and (\ref{wsben}) for small
$U_b/s$ is shown in Fig.~\ref{piVb}, and we observe a good match when
$U_b/s$ is small. For large $U_b/s$ and $U_d/s$, a fit to the
numerical data shows that the fixation probability decreases linearly
with $U_b$.

%-------------------------------------------------------------------
{\subsection*{Evolution of mutation rates in finite populations}

The drift-barrier hypothesis states that in a finite population, the
beneficial effect of lower deleterious mutation rate can be outweighed by the stochastic effects of
random genetic drift which 
limits the evolution of mutation rates \citep{Lynch:2010b}. In a finite population of size $N$, a mutation that decreases the deleterious mutation rate confers an indirect selective advantage and
will spread through the population. However, as $U_d$ decreases, the fixation probability of such a mutant decreases until it reaches
its neutral value $\pi_{\textrm{neu}}=1/N$. Here we have calculated the fixation probability $\pi_0$ 
neglecting stochastic fluctuations. The full {fixation probability} $\Pi$ that includes the neutral and the large population limit may be obtained as follows.} 

The fixation time for a mutator in a finite population of
nonmutators when all mutations are
deleterious has been
calculated using a diffusion theory by \citet{Jain:2013}, and shown to increase exponentially with the population
size. The fixation probability $\sim e^{-2 N S}$ is thus
exponentially small in the population size \citep{Kimura:1980,Assaf:2011},  where we have
identified  the rate of decrease of fixation probability with a 
selection coefficient $2 S$. This effective selection coefficient is
found to match exactly with the result
(\ref{Pi0analy}) for the fixation probability $\pi_0$ obtained here using a
branching process. Although 
this is not a rigorous proof, these observations 
strongly suggest that the fixation probability of a nonmutator in a
finite population of size $N$ is of the classical form \citep{Kimura:1962}
\be
\Pi= \frac{1-e^{-2 S}}{1-e^{-2 N S}}~,~
\label{Kimura}
\ee 
where
$S=\pi_0/2$. We also mention that
the probability $2 S$ depends on the difference in
the deleterious mutation rate of the mutator and the nonmutator when
the mutation rate of the nonmutator is nonzero
\citep{Jain:2013}, and has also been shown to be 
insensitive to the distribution of selective effects \citep{Desai:2011}. 

\begin{figure} 
\centering
\includegraphics[width=0.7\textwidth,angle=-90]{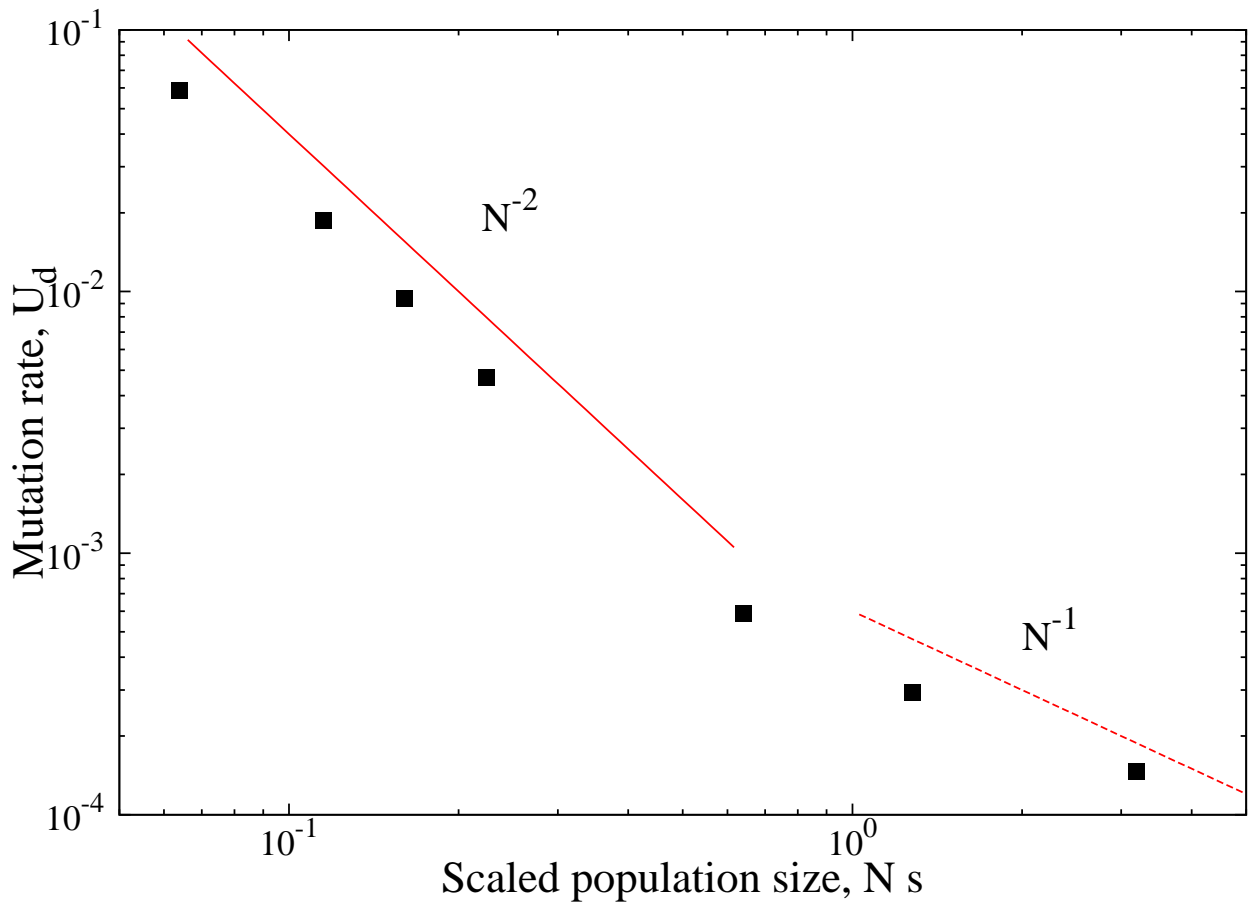}
\label{mut1}
\caption{Relationship between the deleterious mutation rate and the
  population size for selection coefficient $s=0.0064$ when
  compensatory mutations are absent. The points are obtained by numerical
  simulations of a Wright-Fisher process, and the lines show the 
  $N$-dependence in (\ref{crossover}).}
\label{mutfig}
\end{figure}

Thus, according to (\ref{Kimura}), a crossover between positive selection and neutral regime occurs when
$\pi_0 \sim N^{-1}$ and gives a lower bound on the mutation rates. 
We recall that the fixation probability $\pi_0$ in (\ref{Pi0analy}) shows a
  transition when $U_d \sim s$, and at this mutation rate, the fixation probability $\pi_0 \sim s$. This translates into a change in the behavior of $U_d$ when $N s$ crosses one, and we have 
\be
{U_d \sim }
\begin{cases}
(s N^2)^{-1}    ~,~ N s \ll 1 \\
N^{-1} ~,~N s \gg 1~.
\end{cases}
\label{crossover}
\ee
Thus in the weak selection regime ($N s \ll 1$), the deleterious mutation rate depends on the selection coefficient, and decreases
faster than when the selection is strong. Figure
  \ref{mutfig} shows the preliminary results of our numerical
  simulations for a finite size population of mutators with mutation rate $U_d$ in which
nonmutators with mutation rate $U_d/2$ can arise with a
certain probability. This population of nonmutators and
mutators evolves via standard Wright-Fisher dynamics, and the time to fix the
nonmutators is measured  \citep{Jain:2013}. For a fixed $N$, the
fixation time is found to 
increase as the mutation rate of the mutator is decreased until a
minimum mutation rate is reached below which the fixation time remains constant. This lower bound, shown in Fig.~\ref{mutfig}, exhibits different scaling behavior in the weak and strong selection regimes, in accordance with (\ref{crossover}).

%=======================================================================
%=======================================================================
%DISCUSSION
%=======================================================================
%=======================================================================
\section*{Discussion}

{\bf Fixation probability:} A rare mutator arising in a population of
nonmutators carries a higher load of deleterious mutations but offers
indirect benefit by producing more beneficial mutations. The fixation
probability of a rare mutator in a finite nonmutator population has been
studied by \citet{Andre:2006} and \citet{Wylie:2009} analytically, and found to vary nonmonotonically with the mutation rate of the mutator. It has been shown that the fixation probability is of the classical form (\ref{Kimura}) where the effective selection coefficient $S$ when scaled by the selective advantage $s$ increases (decreases) when the ratio of mutation rate to
selection coefficient is below (above) one. Here, we studied a situation in which a
nonmutator appears in a mutator population and is beneficial since it
produces fewer deleterious mutations, and calculated its fixation
probability $\pi_{\textrm{tot}}$ using a branching
process.  The mutator population is assumed to be at
mutation-selection balance and therefore, by definition, selective
sweeps resulting in the spread of favorable mutations are
neglected. However, it is interesting to note that the scaled fixation probability of the nonmutator obtained here also changes its behavior when the deleterious mutation rate is of the order of the selection coefficient, see (\ref{Pi0analy}). 
Our work significantly extends the previous result of \citet{Lynch:2011} as
the deleterious
effect of mutations is allowed to be mild here, and therefore we are
dealing with a truly multilocus problem.

  Compensatory mutations that alleviate the effect of
    deleterious mutations are found to have a surprising effect on the fixation
    probability of the nonmutator.  Although they improve 
  the fitness of the mutator population, it also means that
  the nonmutator can arise in a better genetic background where it has a better
  chance of fixation. Thus, compensatory
  mutations affect both the resident mutator population and the
  invading nonmutator allele in a positive manner. The effect of 
  these two factors on the fixation probability of the nonmutator is, however, 
  opposite and can 
  result in an unexpected increase in the fixation probability of the
  nonmutator 
  when compensatory mutations are present. 
  Here we have shown analytically that this scenario is realised when the mutations are
  weakly deleterious and the compensatory mutation rate is small, as illustrated in Fig.~\ref{piVb}. 
The increase in the fixation probability due to compensatory mutations can be quite high, but we do not have analytical estimates for
this. An exact solution of (\ref{Qdyn}) would, of
course, pave the way for a better analytical understanding 
but is currently not available.

{\bf Fixation time:} In a maladapted  
asexual population, the mutators can sweep the population since they
facilitate rapid 
adaptation \citep{Raynes:2014}. But as the population adapts and the 
supply of beneficial mutations diminishes, mutators have a detrimental
effect on the  
population fitness and a mutation that lowers the mutation rate is
favored.  
In bacteria {\it E. coli}, several genes (such as
${\it mut T}$ and ${\it mut Y}$)  
are involved in avoiding or repairing the errors that occur during the
replication process, and defects in these genes can lead to the
mutator phenotype \citep{Miller:1996}. But compensatory mutations in 
the defective error-repair machinery can reduce the mutation rate, at
least, partially \citep{Wielgoss:2013}. We therefore model this
situation by assigning a probability $b$ with which mutators can  
convert into nonmutators due to a mutation in the
proofreading or error-repair region.  In ${\it E. coli}$, the
conversion probability $f$ from nonmutator to 
mutators has been estimated to be $\sim 10^{-6}$ per bacterium per
generation \citep{Boe:2000}. But the probability $b$ for the reverse
mutation is not known, although one expects $b < f$, possibly
  because it is a gain-of-function mutation \citep{Wielgoss:2013}.

When the rate $N b$ at which the nonmutators are produced from the
mutators is small enough that the new alleles behave independently, the time taken to 
fix the nonmutator population is given by $T= (N b \pi_{\textrm{tot}})^{-1}$. In a long term evolution experiment on $E.coli$, \citet{Wielgoss:2013} 
found the mutation rate to decrease by about a factor two in a nearly
adapted mutator population with a mutation rate $150$ times that of
the wildtype in two lineages. As the population size in Lenski's
experiments has been estimated to be about $10^{7}$
\citep{Wahl:2002}, the product $N b$ can be at most ten which is not too
large. We first note that in the experiment of \citet{Wielgoss:2013}, the fixation 
time was longer in the lineage 
in which the mutation rate decreased by a smaller amount, in accordance
with (\ref{Pi0analy}). To make a quantitative comparison, we
consider the ratio of
the times for the two lineages, as $T$ depends strongly 
on the probability $b$ which is not known experimentally. Using the data in Table 2 of \citet{Wielgoss:2013}, we find the ratio of fixation time in {\it mutT
  mutY-L} background to that in {\it mutT mutY-E} background to be
$9209/5157 \approx 1.8$. The theoretical formula (\ref{Pi0analy}), on
replacing $U_d$ by the difference between the mutation rate of the
nonmutator and mutator 
yields $1.5$ ($1.2$) when mutations are
assumed to be strongly (weakly) deleterious and the selection coefficient 
same in both lineages.  
Since (\ref{Pi0analy}) is obtained assuming that the mutators are
strong whereas the mutation rates decreased merely by a factor two in
the experiment, a more careful examination is 
needed. Solving (\ref{Piktfull}) numerically in the stationary state,
we find that the ratio is unaffected when the mutations are strongly
deleterious. But using the mutation rates in Table 2 of
\citet{Wielgoss:2013} and $s \sim 0.01$ yield the ratio to
be about $4.5$. Although the theoretical conclusions ($1.5-4.5$) are
in reasonable agreement with experiments, the above analysis suggests that the reversion
probability $b$ may not be too small ({\it i.e.} $N b \gtrsim 1$), and a
more sophisticated theory that takes care of the interference between
the nonmutators \citep{Gerrish:1998} may be required to obtain a
closer match. 
We close this discussion by noting that in an experiment on {\it
  S. cerevisiae} in which the adapted population reduced its
genome-wide mutation rate by almost a factor four in two of
the experimental lines \citep{Mcdonald:2012}, the fixation time seems to increase with the mutation rate, in
contradiction with the experiment of \citet{Wielgoss:2013} and the theory presented here. 

{\bf Evolution of mutation rates:} Experiments show that the mutation rate decays as $N^{-0.7}$ for prokaryotes and
$N^{-0.9}$ for eukaryotes \citep{Sung:2012}. 
The population size and deleterious mutation rates are negatively
correlated since deleterious mutations can get fixed in small populations due to
stochastic fluctuations, but not in large populations where the
genetic drift is ineffective \citep{Lynch:2010b}. 
Here, we have shown that a reciprocal relationship between the
population size and mutation rate holds for large populations, but for
small populations, the deleterious mutation rate decreases much
faster, see Fig.~\ref{mutfig}. This is in contrast to experimental
results mentioned above where the data has been fitted assuming a {\it
  single} scaling law. In view of our theoretical results discussed above, a more careful analysis of experimental
data  is required.

While the evolution of deleterious mutation rate has received much
attention, to the best of our knowledge, analogous theoretical predictions for the beneficial mutation rate 
are not available. As large populations experience clonal
interference \citep{Gerrish:1998} which results in the wastage of beneficial mutations, the
rate of beneficial mutations is observed to be smaller in large 
populations in microbial experiments \citep{Perfeito:2007}. An understanding of the relationship between the population size and the rate of beneficial mutations would be an interesting avenue to explore. 
Other potential factors that can affect the correlation between the mutation rate and the population size include epistasis and recombination. Here, we have also ignored the cost of fidelity, and it
remains to be seen how the results presented here are 
affected on including it \citep{Kimura:1967,Kondrashov:1995,Dawson:1998}. 
A more detailed understanding of the mutation rates, both empirically and theoretically, remains a goal for the future.

\vspace{1 cm}

{\bf Acknowledgements:} We thank S. John for help with the numerics, 
and B. Charlesworth and A. Nagar for
helpful discussions. A. James would like to thank CSIR for the
funding. We also thank two anonymous referees and the Associate Editor
for many helpful suggestions and comments. 

\clearpage

%=======================================================================
%=======================================================================
%APPENDIX
%=======================================================================
%=======================================================================
\appendix
\numberwithin{equation}{section}

%---------------------------------------------------------------------------
%---------------------------------------------------------------------------
%APPENDIX 1
%---------------------------------------------------------------------------
%---------------------------------------------------------------------------
\section{Mutator frequency when compensatory mutations are included}
\label{apdetda}

For small selection coefficient and mutation rates, the mutator frequency $p(k,t)$ obeys the following continuous time equations: 
\begin{equation}\label{Qdyn}
  \begin{split}
\frac{\partial p(0,t)}{\partial t} &= -U_d p(0,t) +U_b p(1,t)+ s
     {\bar k} p(0,t) \\
\frac{\partial p(k,t)}{\partial t} &= -U p(k,t) +U_d p(k-1,t)+U_b p(k+1,t)- s
(k-{\bar k}(t)) p(k,t)~,
\end{split}
\end{equation}
where $U=U_d+U_b$ and ${\bar k}(t)=\sum_{k=0}^\infty k ~p(k,t)$. In
the stationary state, the LHS is zero and the frequencies are
time-independent. On dividing both sides of the above equations by $s$, we find that the stationary frequency $p(k)$ depends on the ratios $U_b/s$ and $U_d/s$. We first expand the fraction $p(k)$ in a power series about $U_b/s=0$ as 
\be
p(k)=\sum_{n=0}^\infty \left( \frac{U_b}{s} \right)^n ~p_n(k) ~,
\label{pertdef}
\ee
where $p_n(k)$ is proportional to the $n$th derivative of $p(k)$ with
respect to $U_b/s$ evaluated at $U_b=0$. The lowest order term $p_0(k)$
is the solution of the steady state of (\ref{Qdyn}) in the absence of
compensatory mutations, and
is known to 
be a Poisson distribution with mean ${\bar k}_0=U_d/s$ \citep{Haigh:1978}: 
\be
p_0(k)=e^{-{\bar k}_0} ~ \frac{{\bar k}_0^k}{k!}~,~k=0,1,...
\label{app_p0k}
\ee
To find the
solution with nonzero $U_b$, we first set the LHS of
(\ref{Qdyn}) equal to zero, and substitute
(\ref{pertdef}) in these equations. On neglecting the terms of order
$(U_b/s)^2$ and higher, we obtain the following equations for
$p_1(k)$: 
\bea
 {\bar k}_1 p_0(0) &=&
-p_0(1) \label{Q0ss} \\
{\bar k}_0 p_1(k-1)-k p_1(k)+  {\bar k}_1
p_0(k) &=& p_0(k)-p_0(k+1) ~,~ k=1,2,...
\label{Q1ss}
\eea
where ${\bar k}_1=\sum_{k=0}^\infty  k p_1(k)$. 
Equation (\ref{Q0ss}) above immediately yields ${\bar
  k}_1=-{\bar k}_0$. Thus, as  expected, the effect of compensatory 
mutations is to decrease the deleterious mutations in a
population. Using this result in (\ref{Q1ss}), after some
simple algebra, we get the following one-term recursion equation for
$p_1(k)~,~k \geq 1$:
\be
p_1(k)=\frac{{\bar k}_0}{k}~p_1(k-1)- \left( \frac{1}{k}+
\frac{{\bar k}_0}{k+1} \right)~p_0(k)~,
\ee
which can be iterated easily to give
\be
p_1(k)= \frac{{\bar k}_0^k}{k!}~p_1(0) - p_0(k)~\left[{\bar k}_0
  (H_{k+1}-1)+H_k\right] ~,
\label{Q1corr}
\ee
where the harmonic number $H_k=\sum_{i=1}^k i^{-1}$ and the fraction
$p_1(0)$ is determined using the
normalisation condition, viz., $\sum_{k=0}^\infty p(k)=1$. Since the
fraction $p_0(k)$ already satisfies this condition, we have the
constraint $\sum_{k=0}^\infty p_1(k)=0$, on using which, $p_1(0)$ can
be found. For large $k$, using 
$H_k \approx \ln k$ in (\ref{Q1corr}), we obtain the expression (\ref{Q1simple}). 
%---------------------------------------------------------------------------
%---------------------------------------------------------------------------
%APPENDIX 0
%---------------------------------------------------------------------------
%---------------------------------------------------------------------------
\section{Fixation probability in the absence of compensatory mutations}
\label{apdet0}

To find the total fixation probability given by (\ref{Ptot}),
  we use the expression (\ref{Pi01}) for the fixation probability and
  (\ref{app_p0k}) for the mutator fraction $p_0(k)$ which is a Poisson
  distribution with mean ${\bar k}_0$. 
When ${\bar k}_0 \ll 1$, we have $\pi_0 \approx p_0(0) \pi_0(0)=2 U_d$. But for ${\bar k}_0
\gg 1$, on summing over the mutator backgrounds in which a nonmutator
can arise, we obtain the total fixation probability to
be 
\be
\pi_0= \sum_{k=0}^{\lfloor {\bar k}_0 \rfloor} p_0(k) \pi_0(k)= 2 s ~\frac{e^{-\lfloor {\bar k}_0 \rfloor}
  \lfloor {\bar k}_0 \rfloor^{\lfloor {\bar k}_0 \rfloor+1}}{\lfloor {\bar k}_0 \rfloor!}~.
\label{Pi0an}
\ee
On using the Stirling's formula $x! \approx \sqrt{2 \pi x} (x/e)^x$ for
large $x$ in the last expression, we immediately obtain
(\ref{Pi0analy}). 
Another way of seeing the result in the large ${\bar k_0}$ regime is by approximating the Poisson-distributed $p_0(k)$ by a Gaussian with mean and variance equal to ${\bar k_0}$ and thus obtain
\bea
\pi_0 \sim 2 s \int_{{{\bar k_0}}-\sqrt{{\bar k_0}}}^{{\bar k_0}} dk ({\bar k_0}-k) ~\frac{1}{\sqrt{{\bar k_0}}} ~e^{- \frac{(k-{\bar k_0})^2}{2 {\bar k_0}}} \sim 2 s \sqrt{{\bar k_0}} \int_0^1 dx~x e^{-x^2}~
 \eea
where we have used 
the fact that the mutator frequency is substantial in the
fitness classes lying within a distance $\sqrt{\bar k_0}$ of the
mean. 

%---------------------------------------------------------------------------
%---------------------------------------------------------------------------
%APPENDIX 2
%---------------------------------------------------------------------------
%---------------------------------------------------------------------------
\section{Fixation probability when compensatory mutations are included} 
\label{apdetdc}

Inserting $\pi_{tot}=\pi_0+(U_b/s) \pi_1$ and ${\overline W}={\overline W}_0+(U_b/s)
{\overline W}_1$ in (\ref{Piss2}), and using the exact equation for
$\pi_0(k)$, we get a rather involved expression for 
$\pi_1(k)$ given by 
\bea
\pi_1(k) = -\frac{{\overline W}_1}{{\overline W}_0}~ \frac{W(k) \pi_0(k)
  (1-\pi_0(k))}{{\overline W}_0-W(k) (1-\pi_0(k))} ~.
\eea
Since all the parameters are smaller than one, we work with the
approximate expression (\ref{Pi01}) for the 
probability $\pi_0(k)$ and arrive at (\ref{Pi1simple}). 

We now calculate the contribution $\pi_1$ given by (\ref{Pi1expr}) when
${\bar k}_0 \gg 1$ using the
expression (\ref{Pi1simple}) for $\pi_1(k)$ and the frequency $p_1(k)$ in
(\ref{Q1simple}). We have 
\bea
\pi_1 &=& -2 s {\bar k}_0 (1-\sum_{{k=\lfloor {\bar k}_0
    \rfloor}}^\infty p_0(k))+2 s \sum_{k=0}^\infty ({\bar k}_0 -k)
p_1(k)- \sum_{{k=\lfloor {\bar k}_0 \rfloor}}^\infty p_1(k) \pi_0(k) \\
&=& 2 s {\bar k}_0 \sum_{{k=\lfloor {\bar k}_0 \rfloor}}^\infty p_0(k) - 2 \sum_{{k=\lfloor {\bar k}_0 \rfloor}}^\infty
({\bar k}_0-k) ~s p_1(k)~,~ 
\label{Pi1int}
\eea
where we have assumed that $U_b, U_d, s$ are
small, but $U_b/s$ and $U_d/s$ are finite. The last expression is
obtained on using the normalisation condition $\sum_{k=0}^\infty
p_1(k)=0$ and the expression for the average ${\bar k}_1$. 
For large ${\bar k}_0$, we approximate
the Poisson distribution $p_0(k)$ by a Gaussian as
\be
p_0(k) \approx \frac{1}{\sqrt{2 \pi {\bar k}_0}}~ e^{-\frac{(k-{\bar k}_0)^2}{2
      {\bar k}_0}}~\left(1- \frac{k-{\bar k}_0}{2 {\bar k}_0} + {\cal O}({\bar k}_0^{-2})\right)~.
\ee
Approximating the sums in (\ref{Pi1int}) by integrals, we finally have
\be
\frac{\pi_1}{s} \approx \left[ {\bar k}_0 -\sqrt{\frac{{\bar k}_0}{2 \pi}} \right] -\frac{2
  {\bar k}_0}{\sqrt{\pi}} ~\int_0^\infty dz \frac{e^{-z^2}}{1+ z
  \sqrt{\frac{2}{{\bar k}_0}}} \approx \sqrt{\frac{{\bar k}_0}{2 \pi}}
\label{Vbres}~,
\ee
where we have carried out an integration by parts in the second
integral on the RHS and neglected subleading terms in ${\bar k}_0$.

\clearpage
%%%%%%%%%%%%%%%%%%%%%%%%%%%%%%%%%%%%%%%%%%%%%%%%%%%%%%%%%%%
%BIBLIOGRPAPHY
%%%%%%%%%%%%%%%%%%%%%%%%%%%%%%%%%%%%%%%%%%%%%%%%%%%%%%%%%%%

\clearpage
%=======================================================================
%FIGURES 
%=======================================================================

\clearpage

\clearpage

\clearpage 

%=======================================================================
%=======================================================================

\end{document}